\documentclass[twocolumn,aps,prl,showpacs, amsmath,amssymb, superscriptaddress]{revtex4}

\usepackage{graphicx}
\usepackage{dcolumn}
\usepackage{bm}

\begin{document}
\title{Dynamics of drying in 3D porous media}
\author{Lei Xu}
\affiliation{HSEAS and Department of Physics, Harvard University,
Cambridge, MA 02138, U.S.A.}
\author{Simon Davies}
\affiliation{ICI Strategic Technology Group, R320 Wilton Centre,
Redcar TS104RF, U.K.}
\author{Andrew B. Schofield}
\affiliation{The School of Physics, University of Edinburgh,
Edinburgh EH9 3JZ, U.K.}
\author{David A. Weitz}
\affiliation{HSEAS and Department of Physics, Harvard University,
Cambridge, MA 02138, U.S.A.}

\date{\today}
\begin{abstract}
The drying dynamics in three dimensional porous media are studied
with confocal microscopy. We observe abrupt air invasions in size
from single particle to hundreds of particles. We show that these
result from the strong flow from menisci in large pores to menisci
in small pores during drying. This flow causes air invasions to
start in large menisci and subsequently spread throughout the entire
system. We measure the size and structure of the air invasions and
show that they are in accord with invasion percolation. By varying
the particle size and contact angle we unambiguously demonstrate
that capillary pressure dominates the drying process.
\end{abstract}

\maketitle

Drying is a ubiquitous natural process; it is of particular
importance and interest for porous media where there are many
important applications, including soil drying in agriculture,
recovery of volatile hydrocarbons from underground oil reservoirs,
spray drying in food and pharmaceutical industries, and drying of
paint and ceramic powders. Drying in porous media is often described
theoretically by invasion percolation(IP)\cite{Wilkinson,Page,Shaw,
Prat} and modeled by immiscible displacement experiments in
2D\cite{Tsimpanogiannis}. However, despite its great importance,
direct imaging of drying in three dimensional(3D) porous media is
rare, largely due to the difficulty of visualization in 3D porous
media. Instead, indirect measurements such as light scattering and
acoustic methods\cite{Page}, conductance measurement\cite{Thompson},
or pressure measurement\cite{Knut,Furuberg,Aker} have been used.
While valuable information is deduced with these techniques, many
important dynamical features of drying, such as the flow pattern,
the structure of the abrupt bursts and the liquid redistribution,
remain poorly characterized. To study them, the critical
investigation by direct imaging is indispensable.

In this paper, we study drying dynamics in 3D porous media by direct
imaging with confocal microscopy. We observe a strong flow from the
menisci in large pores to the menisci in small pores during drying.
The flow causes abrupt air invasions starting from large menisci
which subsequently spread throughout the entire system. The size,
structure and dynamics of these invasions are explained by 3D
invasion percolation coupled with liquid
redistribution\cite{Knut,Furuberg}. The liquid redistribution is
confirmed by direct observation. By varying the particle size and
contact angle we also demonstrate unambiguously that capillary
pressure dominates the drying process.

Porous media are prepared by evaporating a high concentration
colloidal suspension, composed of fluorescently labeled PMMA
particles(density $\rho_{pmma}=1.19g/cm^3$ and index of refraction
$n_{pmma}=1.49$) suspended in decalin($\rho_{dec}=0.897g/cm^3,
n_{dec}=1.48$). The particles and solvent are closely index matched,
allowing 3D visualization through more than one hundred particle
layers using confocal microscopy. Samples are dried on clean glass
substrates with the initial volume fractions between $30\%$ and
$40\%$.

The overall drying process is demonstrated by evaporating a drop of
suspension($3.8\mu l$) with particle size $d=1.1\pm0.1\mu m$. The
drying is divided into two stages:  The suspension is initially
compacted as decalin evaporates and the colloid volume fraction
increases. Due to the coffee-ring effect \cite{Deegan}, the
suspension compacts at the droplet edge with the
randomly-close-packed particles, creating a network of pores. Images
of horizontal(xy) and vertical(xz) cross sections taken at the
interface between the packed region and free suspension are shown in
Fig.1(b). Particles can not move once they are in the packed region.
Interestingly, the xz section indicates that the packed region forms
a ``ceiling'' on the top of the droplet due to the lowering of the
air-liquid interface from evaporation. As solvent evaporates, the
dense region at the edge grows towards the center and eventually
fills the entire system (Fig.1(a), top three images). In the second
stage, air invades the system. Cracks appear first, indicating large
stresses in the system, then air continues to invade the network of
pores until all the liquid is evaporated(bottom three images).  It
is the second stage which is the focus of this study.

\begin{figure}[!htp]
\begin{center}
\includegraphics[width=3.2in]{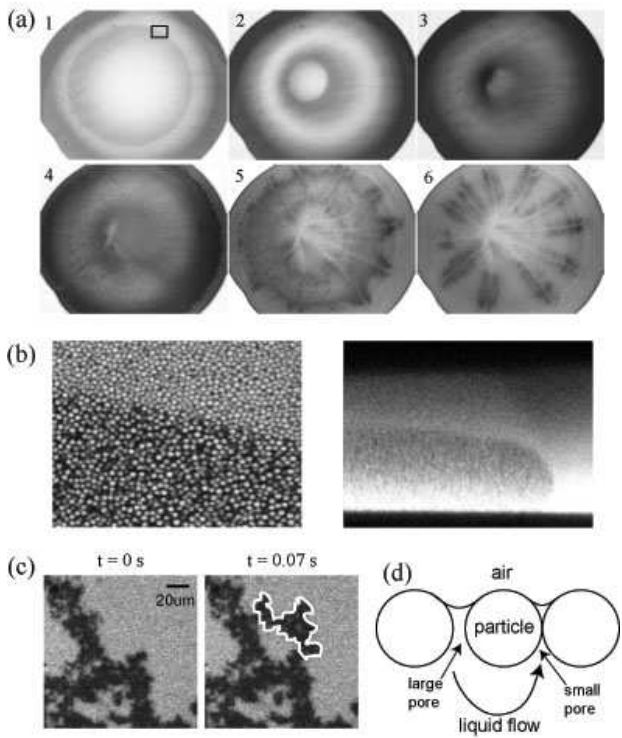}%
\caption{Photographs of drying for $d=1.1\pm0.1\mu m$ colloidal
suspension. (a), drying images taken from above at six different
times: 0s(pict.1), 270s(pict.2), 300s(pict.3), 320s(pict.4),
365s(pict.5) and 410s(pict.6). We divide the process into two
stages: compacting stage(top row) and air invading stage(bottom
row). (b), interface between the packed region and free suspension,
such as the small box area in pict.1. of (a). The left picture is a
horizontal slice and the right is a vertical slice. In the right
picture, a ``ceiling'' of closely packed particles near the
air-liquid interface is clearly visible. (c), image of one large
burst. An area of 800 particle size (the highlighted area) is
invaded within $0.07s$. (d), cartoon illustrating the flow from
large pore to small pore during drying.}
\end{center}
\end{figure}

This second stage is characterized by a series of abrupt air
invasions, or bursts\cite{Haines,Morrow,Knut,Furuberg,Aker}. A
typical burst is identified with confocal microscopy by a bright
area suddenly turning black, corresponding to the displacement of
decalin by air, which is optically mismatched, as illustrated by the
highlighted area in Fig.1(c). In a single step, within 0.07s, an
area of about 800 particle area is invaded. This abrupt behavior
indicates that some large stress is driving the process.

The dominant stress during drying is the capillary pressure from the
tiny menisci\cite{Dufresne}:
\begin{equation}\label{eq1}
\Delta P = \frac{2\sigma cos\theta}{r}
\end{equation}
\noindent where $\sigma$ is the surface tension coefficient,
$\theta$ is the solvent-particle contact angle and $r$ is the radius
of curvature which is typically $r\simeq 0.1 R$ for randomly close
packed spheres of radius $R$. Since $\sigma_{dec}=31mN/m$ and
$cos\theta \simeq 1$, Eq.(1) gives an enormous pressure: $\Delta P
\sim 6\times 10^5Pa\sim 6atm$ for $d=1.1\mu m$ particles! Relative
to the ambient pressure surrounding the droplet, $P_0$, the liquid
within the pores has a very low pressure: $P_0 - \Delta P$.
Therefore every meniscus acts like a low pressure pump trying to
suck liquid from other places. Due to the inhomogeneity of the pore
sizes, the menisci in small pores can produce lower pressure and
draws liquid from menisci in large pores, as illustrated by the
cartoon in Fig.1(d). We emphasize that this flow is quite different
from the capillary flow in the coffee-ring problem, wherein the
faster evaporation rate at the droplet edge leads to bulk flow
outwards from the center to the edge. This flow would move the
air-liquid interface rapidly through the large pores, creating
bursts. The bursting process will terminate either when all the
pores are small enough to balance the capillary pressure or when the
displaced liquid flows to the nearby menisci and reduces the
capillary pressure. Previous research in 2D systems proposed that
the later mechanism results in an exponential cut-off of burst
size\cite{Knut,Furuberg}.
\begin{figure}[!thbp]
\begin{center}
\includegraphics[width=3.4in]{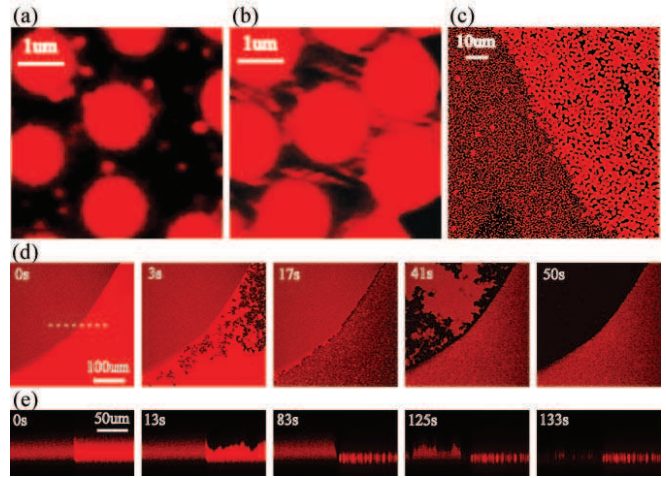}%
\caption{Observation of strong flow from large pores to small pores.
(a), $d=2\mu m$ sample with $d=0.22\mu m$ tracer particles. The
tracer particles move randomly before the invading stage. (b),
tracer particle trajectories show a strong flow during invading
stage. (c), one sample with small pore and large pore domains. The
darker domain on the left is composed by $1.1\mu m$ particles and
the brighter domain on the right is by $2\mu m$ particles. (d), x-y
scans of drying process of the two-domain sample. Large-pore domain
is invaded first. Surprisingly, the invasion is restricted in the
large-pore domain until it has been completely invaded. (e), x-z
sections at a position similar to the dotted line in (d).}
\end{center}
\end{figure}

To visualize this flow, we add a small quantity of tiny tracer
particles($d=0.22\mu m$) to a suspension of larger spheres with
$d=2\mu m$, as shown in Fig.2(a). The tracer particles are small
enough to move freely through the pores. During the compaction
stage, the tracer particles move randomly\cite{footnote1}. However,
upon entering the invasion stage, a strong flow appears, as shown in
Fig.2(b). The flow velocity and direction vary with time and
location, with the magnitude measured from $5$ to $50 \mu m/s$. The
flow is not from center towards edge, thus we exclude the
possibility of the ``coffee-ring'' effect. This provides direct
evidence of the flow in Fig.1(d).

To confirm that the flow is indeed from large to small menisci, we
prepare samples comprised of two distinct domains of differing
particle diameters (d=1.1 and 2 um), illustrated in Fig.2(c).
Various times in the drying process at the boundary of the two
domains are highlighted by the xy sections shown in Fig.2(d). The
large-pore domain (right) is invaded first, proving that air
invasions start in large pores. Strikingly, fluid displacements are
restricted to the large-pore domain until it has been completely
invaded. Since evaporation occurs everywhere, the only way to halt
air invasions at the boundary between the domains is to cause a
macroscopic flow from the large-pore to the small-pore domain to
compensate for evaporation. The large-pore domain effectively serves
as a liquid reservoir which prevents drying of the small-pore domain
until the reservoir is completely exhausted. A similar process in
the xz direction is shown in Fig.2(e). This phenomenon is further
proof confirming the flow pattern from large to small menisci shown
in Fig.1(d). Moreover, it provides a means to precisely control the
order of drying. It also illustrates a method to spontaneously
concentrate material to small-pore regions through drying.

\begin{figure}[!hbp]
\begin{center}
\includegraphics[width=3.2in]{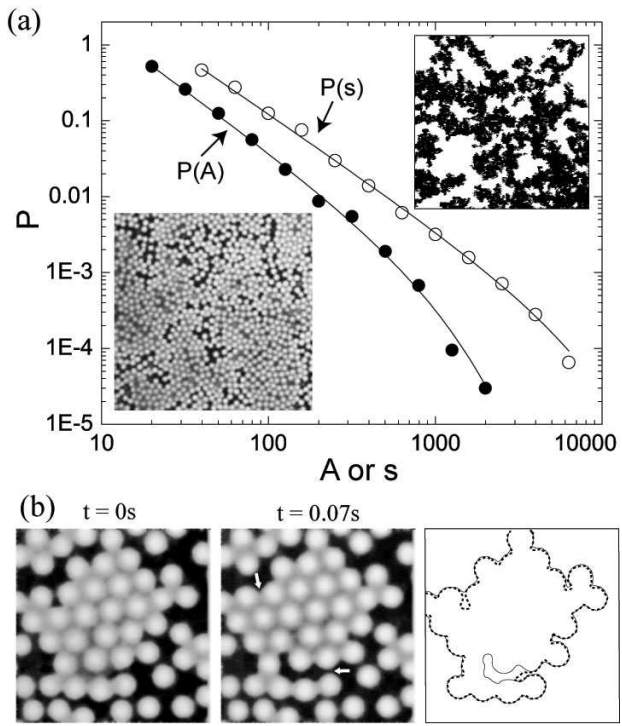}%
\caption{Size, structure and dynamics of bursts. Main panel of (a),
the probability, $P$, of finding a burst of size $s$ or $A$. Both
the 2D ($\bullet$) and 3D ($\circ$) size distributions are plotted.
The 3D size, $s$, is computed from the 2D size, $A$, with the
relationship $s\sim A^{D_f/2}$. The two fitting functions are:
$P(A)\sim A^{-\alpha} Exp(-A/A^*)$, with $\alpha = 1.6\pm0.1$, $A^*
= 850\pm 150\mu m^2$ and $P(s)\sim s^{-\tau'}Exp(-s/s^*)$, with
$\tau'=1.5\pm 0.1$, $s^*=6500\pm1000$. The smallest area in plot is
limited by the image resolution. Lower inset, zoom-in picture of
freshly invaded region. Only $15\%$ pore space is filled by air.
Upper inset, global pattern by many invasions. The field of view is
$345\mu m \times 345 \mu m$. (b), liquid redistribution due to a
burst. The 1st image is before the burst and The 2nd is after. An
invasion occurs at the lower arrow spot, resulting in an obvious
meniscus re-adjustment at the upper arrow spot. The 3rd picture
traces the air-liquid interface of the first two images. The dotted
curve is before the burst and the solid one is after. It shows
menisci re-adjustments at several positions.}
\end{center}
\end{figure}

The flow evacuates the liquid from large pores and leads to abrupt
bursts. The sequence of bursts has been studied indirectly through
the associated pressure changes in similar
systems\cite{Knut,Furuberg}; however, there has been no structural
characterization of these events. We directly visualize these bursts
by horizontally scanning a fixed area over time. The horizontal
scanning requires drying to proceed horizontally as well. We achieve
this by drying the sample between two parallel substrates. We coat
both substrates with a layer of PMMA to match the wetting property
with the bulk particles. Since the invaded air strongly scatters the
light, we restrict our measurements to regions near the bottom
substrate, where the measurement is most reliable. Images are
acquired at $2.5$ frames per second, and the burst areas are
measured by subtracting two adjacent frames (see fig 1(c)).  More
than 600 bursts in a single drying process are measured to obtain
good statistics. We find that the area of the bursts ranges from a
single particle to hundreds of particles. The probability of a
burst, $P$, as a function of its area, $A$, exhibits a power-law
distribution for small areas but rapidly decreases for large areas
as shown by the log-log plot($\bullet$) in the main panel of
Fig.3(a). The power-law distribution is in accord with the invasion
percolation and the deviation at large areas reflects the fact that,
during the bursts, liquid does not have time to evaporate but
instead gets redistributed among the menisci. This redistribution
decreases the capillary pressure and prevents the bursts from
growing too large. The functional form $P(A) \sim A^{-\alpha}
Exp(-A/A^*)$\cite{Knut,Furuberg}, accounts for both the power-law
behavior of invasion percolation and the exponential decay due to
redistribution of the liquid and fits the data well as shown by the
solid line in Fig.3. We obtain $\alpha = 1.6\pm0.1$ and $A^* =
850\pm 150\mu m^2$.

The exponent differs from previous work which found $\alpha=1.32$
\cite{Knut,Furuberg}. However, this was for a 2D network of pores
whereas our system is roughly one hundred particle layers in
thickness and thus more nearly 3D. We examine the freshly invaded
regions\cite{footnote2} at single particle level and find a complex
structure, as shown in the lower inset of Fig.3(a). Instead of being
completely filled by air, only $15\sim25\%$ of the invaded region is
filled by air. This value agrees with 3D bond percolation
thresholds, $12\sim25\%$ \cite{Stauffer}. Scans at different heights
also show that the invaded region is a 3D connected network of air.
These results suggest that each burst is a 3D invasion percolation
event. Therefore we compute the true 3D size of a burst, $s$, from
the 2D area, A: $s\sim A^{D_f/2}$, with $D_f=2.5$ the fractal
dimension of 3D percolation. We plot $P(s)$ in the main panel as
open symbols($\circ$) and fit the data with: $P(s)\sim
s^{-\tau'}Exp(-s/s^*)$, and obtain $\tau'=1.5\pm 0.1$,
$s^*=6500\pm1000$. Theoretical work showed that the power $\tau'$ is
related to percolation exponents\cite{Robbins1}:
$\tau'D_f=D_f+D_e-\nu^{-1}$. This relation predicts $\tau'=1.55$ in
3D and $1.31$ in 2D. Our data are in excellent accord with the 3D
result.

The sequence of bursts quickly spreads throughout the field of view
and forms a macroscopic pattern, as shown in the picture taken at
the moment when invasions first percolate the field of view($345\mu
m \times 345 \mu m$) in the upper inset of Fig.3(a). The pattern is
fractal-like and similar to previous experiment\cite{Shaw}. However,
from the magnified image in lower inset, we determine that the black
region is not completely filled by air, but instead formes 3D
percolated networks of air.

We can also directly determine the origin for the exponential
cut-off: the liquid redistribution. In two frames separated by
$0.07s$, we catch a single burst and the subsequent liquid
redistribution. The two frames are shown in the two pictures of
Fig.3(b) on the left. The first picture is before the burst while
the second is after. The burst occurs at the spot indicated by the
lower arrow and an obvious meniscus re-adjustment occurs at the spot
indicated by the upper arrow. More careful comparison is done in the
third picture of Fig.3(b), where we trace the air-liquid interfaces
of the first two pictures. The dotted line is the interface before
the burst while the solid line is after. There are meniscus
re-adjustments in several places. These re-adjustments reduce the
capillary pressure and lead to the exponential cut-off, $s^*$. This
observation provides unambiguous evidence for the redistribution
argument proposed by previous indirect
measurements\cite{Knut,Furuberg}.

We also vary both the pore size and the contact angle, to further
validate the prediction of Eq.(1). The effect of pore size is
demonstrated by comparing two completely dried samples with
particles of $d = 0.22\mu m$(left) and $d = 2\mu m$(right) in
Fig.4(a). The beautiful crack pattern in the $d = 0.22\mu m$ sample
indicates large stresses during drying, while the absence of cracks
in the $d = 2\mu m$ sample suggests that the stresses are much
lower; This is consistent with Eq.(1). We vary the contact angle of
the invading fluid by studying the re-hydration of a dried sample.
We see a marked difference in re-hydration: the liquid-air interface
sweeps the whole field of view in a smooth and continuous manner, in
sharp contrast with the fractal pattern of drying in Fig.3(a). This
demonstrates the important effect of $\theta$, as predicted in
Eq.(1). This result is in great agreement with a previous
simulation, which explained the smooth interface by the cooperative
activity of neighboring menisci\cite{Robbins2,Robbins3}.

\begin{figure}[!htp]
\begin{center}
\includegraphics[width=3.2in]{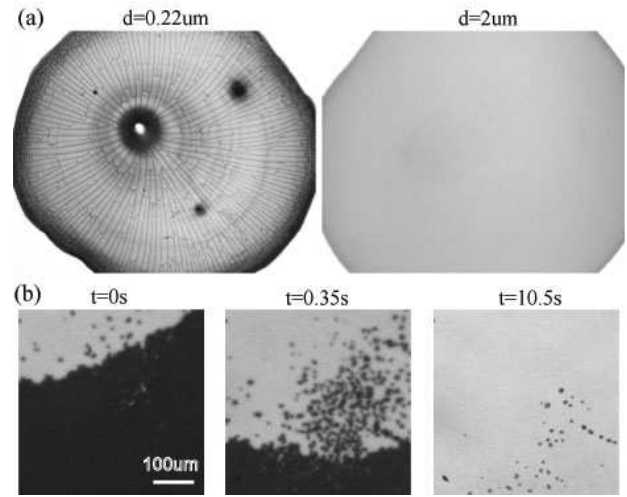}%
\caption{Variation of particle size and contact angle. (a),
completely dried samples of small ($d=0.22\mu m$, left) and large
($d=2\mu m$, right) particles. There are many cracks for $0.22\mu m$
sample and no cracks for $2\mu m$ one. (b), re-hydration of a dried
sample. We see a smooth and continuous interface, in sharp contrast
to the fractal pattern of drying.}
\end{center}
\end{figure}

By using direct imaging, we show that drying of a 3D porous media
can be well described by invasion percolation provided the effects
of liquid redistribution are included; which leads to an exponential
cut-off of the power-law distribution of abrupt invasions. The
liquid redistribution is directly observed, verifying the important
feature that had heretofore only been deduced by indirect methods.
Our study reveals the flow pattern of drying which may ultimately
afford a means to control drying in porous media. A deep
understanding of drying may provide a means to help eliminate
cracking or other undesirable effects.

This work was supported by ICI with partial support from the
NSF(DMR-0602684) and the Harvard MRSEC(DMR-0213805). We thank Mark
Robbins for helpful discussions.


\begin{thebibliography}{99}

\bibitem{Wilkinson} D. Wilkinson and J. F. Willemsen, J. Phys. A
{\bf 16}, 3365 (1983).

\bibitem{Page} J. H. Page \emph{et al}, Phys. Rev. Lett. {\bf 71},
1216 (1993).

\bibitem{Shaw} T. M. Shaw, Phys. Rev. Lett., {\bf 59}, 1671 (1987).

\bibitem{Prat} M. Prat, Inter. Journal of Multiphase Flow, {\bf 19},
691 (1993).

\bibitem{Tsimpanogiannis} I. N. Tsimpanogiannis and Y. C. Yortsos,
Phys. Rev. E, {\bf 59}, 4353 (1999).

\bibitem{Thompson} A. H. Thompson, A. J. Katz and R. A. Raschke,
Phys. Rev. Lett., {\bf 58}, 29 (1987).

\bibitem{Knut} K. J. M\aa l\o y, L. Furuberg, J. Feder and T. J\o
ssang, Phys. Rev. Lett., {\bf 68}, 2161 (1992).

\bibitem{Furuberg} L. Furuberg, K. J. M\aa l\o y and J. Feder, Phys. Rev. E, {\bf
53}, 966 (1996).

\bibitem{Aker} E. Aker \emph{et al}, Europhys. Lett., \textbf{51}, 55
(2000).

\bibitem{Deegan} R. D. Deegan \emph{et al}, Nature {\bf 389}, 827
(1997).

\bibitem{Haines} W. B. Haines, J. Agr. Sci., {\bf 20}, 97 (1930).

\bibitem{Morrow} N. R. Morrow, Ind. Eng. Chem., {\bf 62}, 32 (1970).

\bibitem{Dufresne} E. R. Dufresne\emph{et al }, Phys. Rev. Lett., {\bf
91}, 224501 (2003).

\bibitem{footnote1} There is a weak "coffee ring" flow as well,
but its magnitude is negligible compared with thermal motion.

\bibitem{footnote2} The result is only valid for freshly invaded
region because the fraction of liquid decreases with evaporation.

\bibitem{Stauffer} D. Stauffer, \emph{Introduction to Percolation
Theory}, (Taylor and Francis, London, 1985).

\bibitem{Robbins1} N. Martys, M. O. Robbins and M. Cieplak, Phys. Rev.
B, {\bf 44}, 12294 (1991).


\bibitem{Robbins2} M. Cieplak and M. O. Robbins, Phys. Rev. Lett, {\bf 60},
2042 (1988).

\bibitem{Robbins3} M. Cieplak and M. O. Robbins, Phys. Rev. B, {\bf 41}, 11508
(1990).


\end{thebibliography}
\end{document}